\documentclass[12pt]{iopart}
\usepackage{epsfig,color,ifpdf}
\usepackage{iopams}
\pdfoutput=1
\begin{document}

\title{Noisy continuous--opinion dynamics}

\author{M. Pineda, R. Toral and E. Hern\'andez-Garc\'{\i}a}

\address{IFISC (CSIC-UIB), Instituto de F\'{\i}sica Interdisciplinar
y Sistemas Complejos, Campus Universitat de les Illes Balears,
E-07122 Palma de Mallorca, Spain} \ead{m.pineda@ifisc.uib-csic.es}

\begin{abstract}
We study the Deffuant {\sl et al.} model for continuous--opinion 
dynamics under the influence of noise. In the original version of
this model, individuals meet in random pairwise encounters after
which they compromise or not depending of a confidence parameter.
Free will is introduced in the form of noisy perturbations:
individuals are given the opportunity to change their opinion,
with a given probability, to a randomly selected opinion inside
the whole opinion space. We derive the master equation of this
process. One of the main effects of noise is to
induce an order-disorder transition. In the disordered state the
opinion distribution tends to be uniform, while for the ordered
state a set of well defined opinion groups are formed, although
with some opinion spread inside them. Using a linear
stability analysis we can derive approximate conditions for the
transition between opinion groups and the disordered state. The
master equation analysis is compared with direct Monte-Carlo
simulations. We find that the master equation and the Monte-Carlo
simulations do not always agree due to finite-size induced
fluctuations that we analyze in some detail.
\end{abstract}

 Non-equilibrium processes: Stochastic particle dynamics (Theory).
\maketitle

\section{Introduction}

The application of techniques and tools from nonlinear and
statistical physics to understand the dynamics of opinion changes
in a society has become a topic of interest in recent years
\cite{castellano}. A society can be thought of as a complex system
composed by a large number of interacting individuals with diverse
opinions. These opinions are not necessarily static, but they
evolve due to a variety of internal as well as external factors,
such as the influence of advertising and acquaintances, amongst
others. As a result of this evolution, a consensus opinion could
emerge (a vast majority on individuals adopting a similar
opinion), or the population could fragment into a number of
groups. To analyze the process of opinion formation, several
models inspired from statistical mechanics have been developed. In
those, the opinion held by an individual is a dynamical variable
which evolves by some rules, usually with an important stochastic
ingredient \cite{stauffer}. Models can be divided in two broad
categories: {\sl discrete models} where the opinion can only adopt
a finite set of values \cite{galam,schweitzer,sznajd}, and {\sl
continuous models} where the opinion of an individual is expressed
as a real number in a finite interval
\cite{deffuant1,deffuant2,deffuant3,krause1}. Discrete models are
useful when analyzing cases in which individuals are confronted
with a limited number of options (a political election, for
example) where one is forced to choose amongst a finite set of
parties. Continuous models are more suitable to analyze cases in
which a single issue (legalizing abortion, for example) is being
considered and opinions can vary continuously from ``completely
against" to ``in complete agreement".

A continuous model introduced by Deffuant and collaborators
\cite{deffuant1} has received much attention recently. This model
implements the {\sl bounded confidence} mechanism by which two
individuals only influence the opinion of each other if their
respective opinions differ less than some given amount. In other
words, people holding too distant opinions on an issue will simply
ignore each other and will, hence, keep their original opinions.
It is only through the interaction of not too distant people that
we manage to modify our opinion. This model was in turn inspired
by the Axelrod model for the dissemination of culture
\cite{axelrod} and general threshold models \cite{granovetter} and
has in turn inspired a large number of extensions and
modifications
\cite{redner,deffuant4,stauffer1,kozma,guo,carletti1,carletti2,ben-naim,lorenz1}.

In the Deffuant {\sl et al.} model individuals meet in random
pairwise encounters in a given connectivity network, but the
subsequent evolution is completely deterministic. This leads to
final states in which either {\sl perfect} consensus has been
reached or the population splits in a finite number of groups such
that all individuals in one group have {\sl exactly} the same
opinion. We believe that such uniform states are not very
realistic and some degree of discrepancy must appear within
otherwise well defined groups. In this paper, we introduce an
additional element of randomness in the dynamics. It aims to
represent, certainly in a caricaturist manner, the element of {\sl
free will} present in all human decisions by which we do not
follow blindly the opinion dictated by our relationships. Our aim
is to analyze how the interplay between this {\sl free will} and
the interactions amongst individuals affects group formation in
opinion dynamics. In the language of statistical mechanics, what
we are doing is to add noise to the deterministic dynamics and
analyze which aspects of the model are robust against the
introduction of noise. Noise is introduced by allowing an
individual opinion to change to another randomly chosen value in
the whole opinion space. Under some circumstances this turns out
to be equivalent to allowing each agent to return, at some random
times, to a specific opinion preferred by him.

Our analysis, based upon numerical integrations of the
corresponding master equation as well as Monte-Carlo simulations,
reveals new and interesting phenomenology. There exists a critical
value $m_c$ of the noise intensity, which depends on the
confidence range, such that for noise larger than this value the
system becomes disorganized and group formation does not occur. We
provide a linear stability analysis that reproduces the
order-disorder transition that occurs at $m_c$. For noise smaller
than $m_c$, the steady-state probability distributions in opinion
space broaden with respect to the noiseless case, but still have
large peaks and group formation can be unambiguously defined by
looking at the maxima of the distributions. The group formation
occurs by a series of bifurcations that mimic those that occur in
the noiseless case.

An important aspect of our work, that we want to stress here, is
that the numerical Monte-Carlo simulations do not necessarily
agree with the results of the master equation. This is due to the
inherent finite-size-induced fluctuations that occur in the
simulations. A similar warning is required when one tries to infer
about possible applications of the model to real situations. For
example, it is possible to find regions of bistability where
dynamical transitions between a single group and two groups occur.
These transitions do not occur in the infinite-size thermodynamic
limit taken routinely in most studies. This stresses the role that
a finite size has on the dynamics of social systems \cite{tt2007}.

This paper is organized as follows. Deffuant {\sl et al.} model is
briefly reviewed in Sec. \ref{sec:model}. The main results are
presented in section \ref{sec:noise}, devoted to study this model
in the presence of noise. In Sec. \ref{sec:linear} we use a linear
stability analysis to derive approximately the critical value of
the noise intensity for the formation of opinion groups. Summary
and conclusions are presented in Sec. \ref{sec:conclusions}.

\section{Review of Deffuant \emph{et al.} model}
\label{sec:model}
Let us consider a population with $N$ individuals. We will denote
by $x_n^i$ the number representing the opinion on a given topic
that individual $i$ has at time-step $n$. As mentioned in the
introduction, the opinion is a real variable in a finite interval
and, without loss of generality, we take $x_n^i\in[0,1]$.
Initially, it is assumed that the values $x_0^i$ for $i=1,\dots,N$
are randomly distributed in the interval $[0,1]$. Dynamics is
introduced to reflect that individuals interact, discuss, and
modify their opinions. In the original version of the model
\cite{deffuant1}, at time-step $n$ two individuals, say $i$ and
$j$, are randomly chosen. If their opinions satisfy
$|x_n^i-x_n^j|<\epsilon$, so that they are close enough, they are
modified as:
\begin{equation}
\label{eq:rule}
\begin{array}{lcr}
x_{n+1}^{i}& =&x_{n}^{i}+\mu(x_{n}^{j}-x_{n}^{i}),\cr
x_{n+1}^{j}& =&x_{n}^{j}+\mu(x_{n}^{i}-x_{n}^{j}),
\end{array}
\end{equation}
otherwise they remain unchanged. Whether the opinions have been updated or not, time increases $n\to n+1$. As a consequence of the iteration
of this dynamical rule, the system reaches a static final
configuration which, depending on the values of the parameters
$\epsilon$ and $\mu$, can be a state of full consensus where all
individuals share the same opinion, or of fragmentation with
several opinion groups. It is customary to introduce the time
variable $t=n\Delta t$, where $\Delta t=1/N$, measuring the number
of opinion updates per individual, or number of Monte-Carlo steps
(MCS).

The parameter $\mu$ is restricted to the interval $(0,0.5]$. It
determines the convergence time between individuals as well as the
number of final groups \cite{laguna,porfiri}. For small values of
$\mu$, the individuals slightly change their opinions during
meeting, while for $\mu=0.5$ the interacting individuals fully
compromise and, after the meeting, they share the same opinion. As
in most studies, we will adopt from now on in this paper the value
$\mu=0.5$. The parameter $\epsilon$, which runs from $0$ to $1$,
is the confidence parameter. Starting from uniformly distributed
random values for the initial opinions, the typical realization is
that for large values, $\epsilon \geq 0.5$, the system evolves to
a state of consensus where all individuals share the same opinion
and that, decreasing $\epsilon$, the population splits into
opinion groups separated by distances larger than $\epsilon$.

\begin{figure}
\centering
\epsfig{file=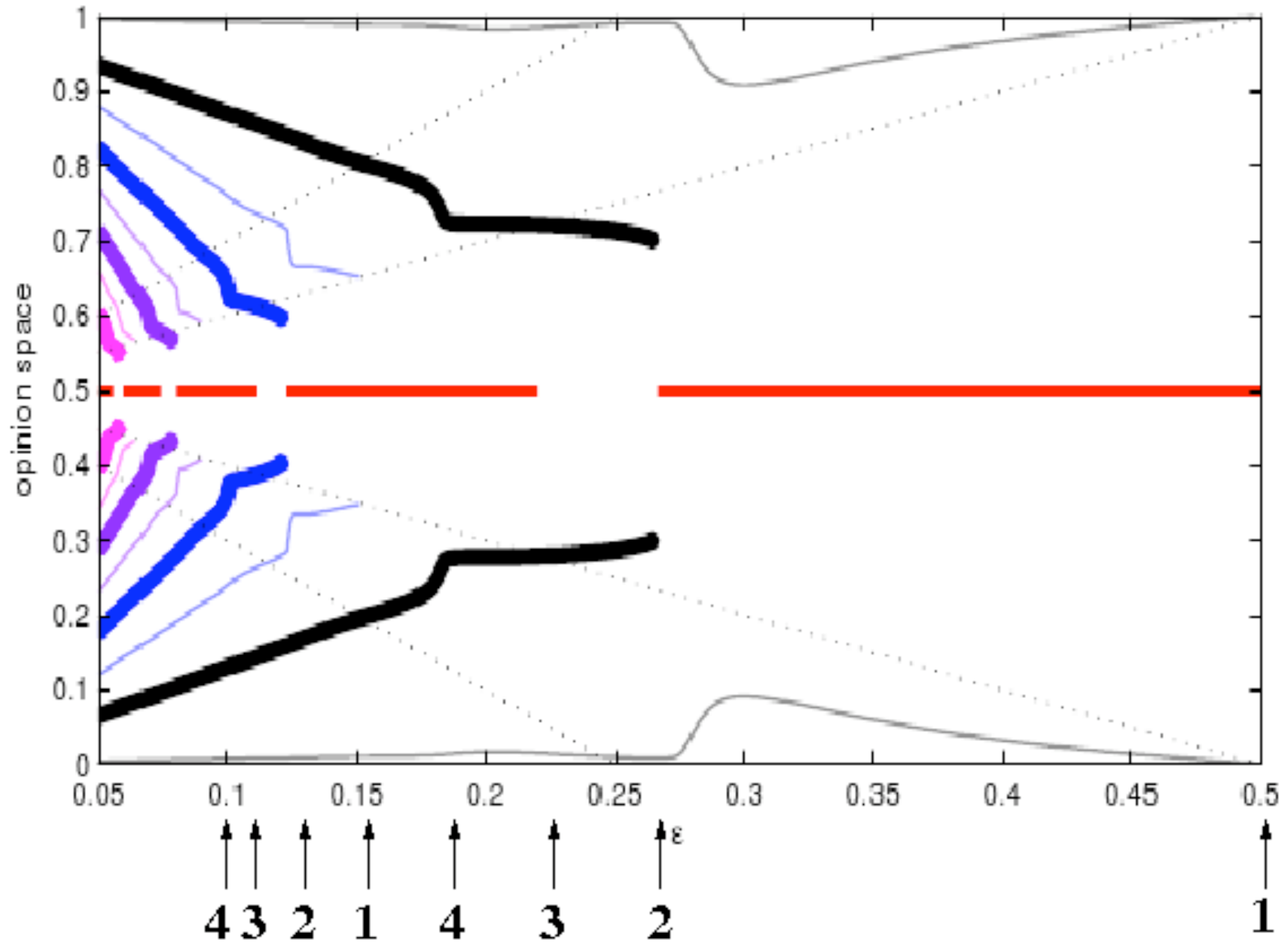,width=0.65\linewidth}\vspace{-1.0truecm}
\epsfig{file=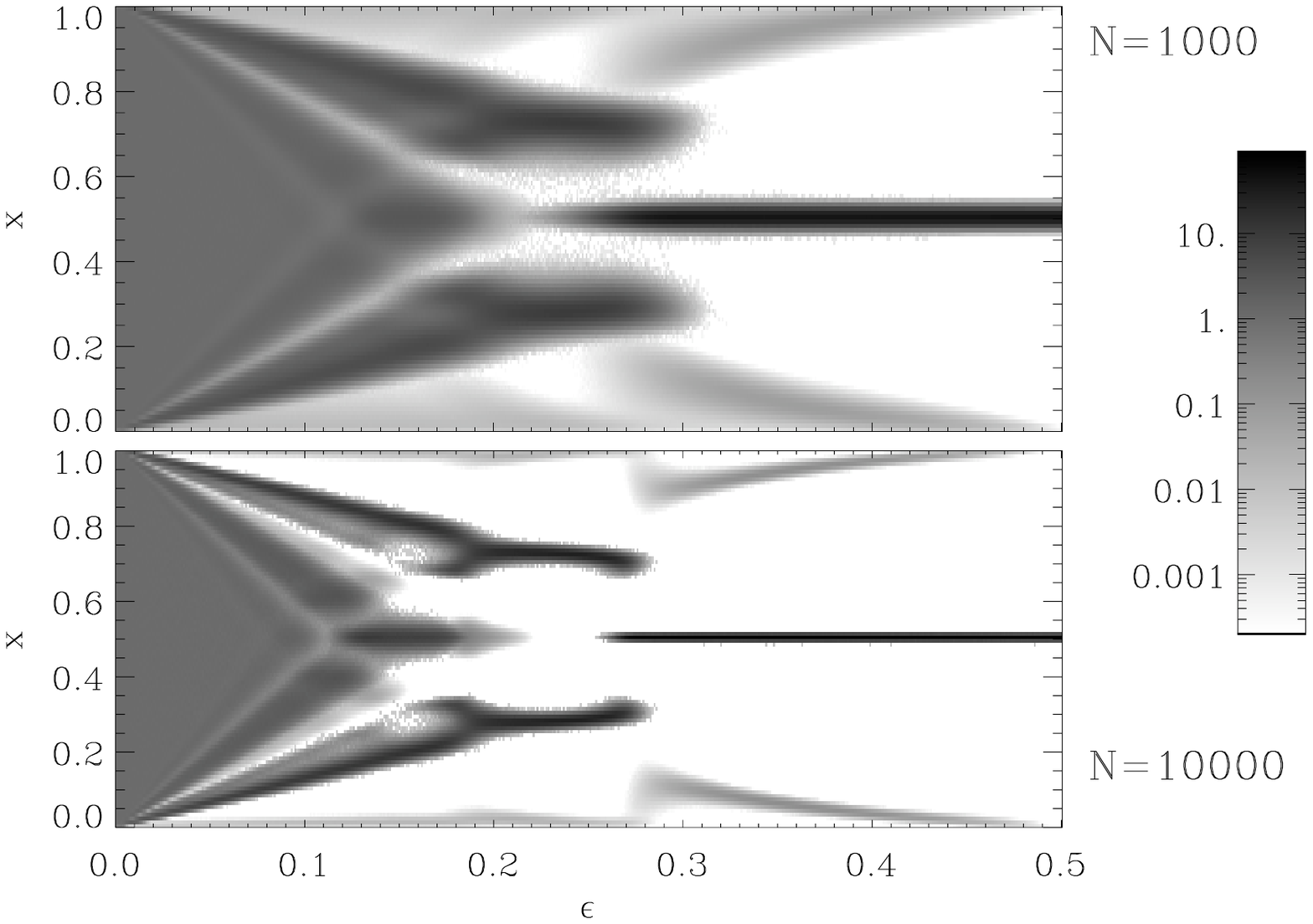,width=0.85\linewidth,height=0.95\linewidth}
\vspace{-6.0 truecm}
\caption{Top panel: bifurcation diagram of the noiseless Deffuant {\sl et al.},
model reproduced from reference \cite{lorenz3}, coming from a
numerical analysis of the master equation of the model
\cite{redner,lorenz2}. Lines show the position of the opinion groups as a function of $\epsilon$. By arrows we indicate the location of the
four basic types of bifurcations: type-1 at $\epsilon\approx0.5$,
type-2 at $\epsilon\approx0.266$, type-3 at $\epsilon\approx0.22$,
type-4 at $\epsilon\approx 0.182$, and this pattern repeating as
$\epsilon$ decreases even further \cite{lorenz2}. In the bottom
panels we plot in a logarithmic grey-scale the asymptotic
probability distribution $ P_{\infty}(x)$ (values smaller than $2\times
10^{-4}$ are plotted white), as a function of $\epsilon$, resulting
from extensive numerical simulations of the microscopic model with
$N=10^{3}$ and $N=10^4$ agents, respectively. Note that a single
realization using the Monte-Carlo microscopic rules of the model
leads to a probability distribution which is a sum of
delta-functions. The distributions displayed in the panels are the
result of an average over $10^5$ realizations for $N=10^3$ and
$2\times10^4$ realizations for $N=10^4$ and a histogram bin size
$\Delta x=0.01$. In all cases, both in the master equation as in
the Monte-Carlo simulations, the initial condition represents opinions
which are random and uniformly distributed in the interval
$[0,1]$.
\label{fig1}}
\end{figure}

The process can be described in terms of a master equation for the probability density function $P(x,t)$ for an individual opinion $x$ at time $t$.
The Appendix contains a derivation of this master
equation in the presence of the additional noise term described in 
section \ref{sec:noise}. Equation (\ref{eq:MEN}) with $m=0$ is the master equation
for the noiseless original Deffuant et al. model, first obtained
in \cite{redner}. A detailed analysis based on its numerical
integration \cite{redner,lorenz1}\footnote{In order to compare
with the results of reference \cite{redner} we note that our
parameter $\epsilon$ is related to their parameter $\Delta$ by
$\epsilon=1/2\Delta$.} shows that there are four basic modes of
group appearance, dominance, or splitting, which are called {\sl
bifurcations} \cite{redner,lorenz1} in this context (see
Fig.~\ref{fig1}): nucleation of two minor groups symmetrically
from the center of the opinion interval (type-1, such as the birth
of two minor groups from the boundaries at $\epsilon=0.5$ in
Fig.~\ref{fig1}); nucleation of two major groups from the central
one (type-2, as occurring at $\epsilon\approx0.266$ in
Fig.~\ref{fig1}); nucleation of a minor central group (type-3, as
occurring at $\epsilon\approx0.222$ in the figure); and, finally,
sudden increase of the mass of that central group accompanied by a
sudden drift outwards in the location of the two major groups
(type-4, at $\epsilon\approx0.182$ in the figure). In this
sequence ``major'' opinion groups contain a high fraction of the
population, while ``minor'' groups contain a much smaller fraction
(of the order of $10^{-2}$ or smaller). The bifurcation pattern
repeats itself as $\epsilon$ decreases even further.

It is important to emphasize that the situation depicted in Fig.
\ref{fig1} is the result for steady solutions of the master
equation attained at long times starting from a uniform initial
distribution. Many other steady solutions of the master equation
exist. In particular, any combination of delta-functions is a
steady solution of the noiseless master equation provided they are
separated by more than a distance $\epsilon$. We stress also that
this analysis based upon the master equation corresponds to the
limit case where the number of individuals $N$ tends to infinity.
In Monte-Carlo simulations of the microscopic rules, or
in practical applications with a necessarily finite number of
individuals, some features need to be considered. It is still true
that each realization ends up in a small number of groups (both of
major and minor type), all individuals within a group holding
exactly the same opinion. However, the exact location of the
groups might vary with respect to the master equation prediction
and minor groups might not appear depending on the particular
realization and the total size of the population. Furthermore,
there could be realizations in which even the number of observed
major groups differs from the prediction of the master equation.
These effects are more pronounced the smaller the number of
individuals. In the same Fig.~\ref{fig1} we have plotted the
distribution of observed groups, averaged over many realizations
for two different number of individuals $N$, where the aforementioned
properties can clearly be observed. For instance, for
$\epsilon=0.28$, the master equation predicts that there should be
only one major group, centered at $x=0.5$, and two minor groups.
However, in almost half of the realizations with $N=1000$
individuals the opinions split instead in two major groups
centered around $x=0.28$ and $x=0.72$ and, eventually, some minor
extreme groups.

\section{Effect of noise}
\label{sec:noise}

Noise is introduced as a random change of an individual's opinion.
Specifically, we modify the dynamics as follows: at time-step $n$
the original dynamical rule, Eq.~(\ref{eq:rule}), applies only
with probability $1-m$. Otherwise, a randomly chosen individual
$i$ changes his opinion to a new value $x_{n+1}^i$ drawn from a
uniform distribution in the interval $[0,1]$ and all other
opinions remain unchanged. The probability $m$ is a measure of the
noise-intensity. Note that, quite generally, this rule is equivalent to allowing each agent to return to a specific, basal, opinion preferred by him, provided that the basal opinions are randomly distributed amongst the agents.
In this section, we will study in detail the
effect of this new ingredient in the dynamical evolution of the
model. We analyze both the results coming from a numerical
integration of the master equation as well as numerical
Monte-Carlo-type simulations of the microscopic rules of the model.

\subsection{Master equation approach}

\begin{figure}
\centering
\vspace{-2.0 truecm}
\epsfig{file=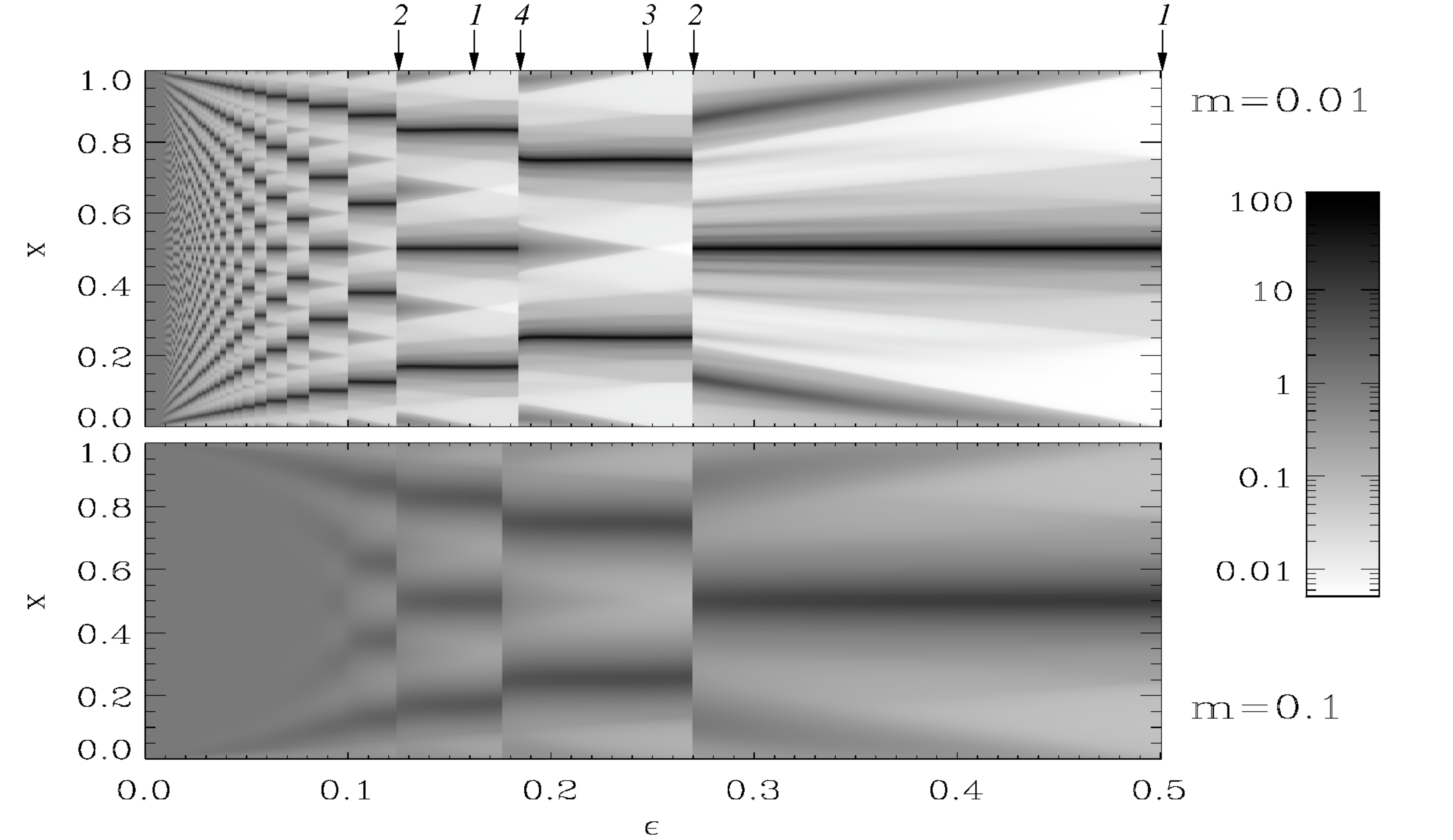,width=1.1\linewidth}
\caption{Plot, in a logarithmic grey scale, of the asymptotic probability distributions
$ P_{\infty}(x)$ as functions of $\epsilon$, obtained after a numerical
integration of the master equation (\ref{eq:MEN}) for $m=0.01$,
top panel, and $m=0.1$, bottom panel. We have used a fourth-order
Runge-Kutta method with a time step $\Delta t=0.1$ for the time
evolution and Simpson's rule for the integrals in $x$-space with a
discretization $\Delta x=1/M$, $M=2000$. We have checked in some
tests cases that smaller time or discretization steps do not
change significantly the results. As in Fig.\ref{fig1} we show by
arrows the location of the bifurcation points.
\label{fig:P_st_master}}
\end{figure}

The Appendix contains a derivation of the master equation, Eq.
(\ref{eq:MEN}), appropriate for this process. We have first
obtained the asymptotic distribution $P_{\infty}(x)=\lim_{t\to\infty} P(x,t)$ of the master equation
starting from a suitable initial condition $P(x,t=0)$. As in other
studies, we assume that the initial condition represents a uniform
distribution in opinion space, i.e. $P(x,t=0)=1$ for $x\in[0,1]$
and $P(x,t=0)=0$ otherwise. For $m=0$ the steady-state
distribution $ P_{\infty}(x)$ it is a sum of delta-functions located at
particular points. In the case $m>0$ the steady distributions are
no longer delta-functions but still are peaked around some
particular values if $\epsilon$ is not too small or $m$ not too
large. We have plotted in Fig.~\ref{fig:P_st_master} the master
equation steady probability distributions $ P_{\infty}(x)$ as a
function of the parameter $\epsilon$ for two different values of
the noise intensity. In the small noise case $m=0.01$ it is still
possible, for not too small $\epsilon$, to identify the same type
of bifurcations than in the noiseless case by looking at the
maxima of the probability distributions: a type-1 bifurcation at
$\epsilon=0.5$ where minor groups begin to form at $x=0$ and
$x=1$; a type-2 bifurcation at $\epsilon\approx 0.2695$ where the
distribution switches from having one single maximum at $x=0.5$ to
having two maxima of equal height located at $x\approx1/4$ and $x\approx3/4$;
a type-3 bifurcation at $\epsilon\approx 0.250$ where a central
maximum begins to grow; a type-4 bifurcation at $\epsilon\approx
0.1835$ where three equally spaced maxima of equal height at
$x\approx 1/6$, $x=1/2$ and $x\approx5/6$ appear. This pattern of
bifurcations repeats as $\epsilon$ decreases even further.
However, type-1 and type-3 bifurcations are somewhat ambiguous
to define since the relative importance of the minor groups
actually increases continuously instead of sharply increasing when
new maxima begin to form. It is worth stressing that, at variance
with the noiseless case, the location of the major groups (defined
as the absolute maxima of the distribution) does not vary with
$\epsilon$ until a new bifurcation of type-2 or type-4 is reached.
These maxima are regularly located at $x\approx 1/k,3/k,\dots, (k-1)/k$
for $k=2,4,6,\dots$

The same general structure can be observed in the case of larger
noise $m=0.1$, although the distributions are much broader now.
The location of the main bifurcation points can be located at
$\epsilon\approx 0.2965$ (type-2) and $\epsilon\approx0.1755$
(type-4). The type-1 and type-3 transitions are very imprecisely
defined, specially for smaller values of $\epsilon$.

For both noise values, one observes that groups become less
defined and finally are replaced by a more or less unstructured
distribution for $\epsilon$ below a critical value $\epsilon_c$
which increases with $m$. Alternatively, one realizes the
existence of a critical value $m_c=m_c(\epsilon)$, increasing with
$\epsilon$, above which the group structure disappears from the
steady distribution.

A somehow expected feature that emerges from the data shown in
Fig. \ref{fig:P_st_master} is that the width of the steady
distributions grows with the noise intensity. An explicit
expression for the width of the single maximum present when
$\epsilon\ge 1$ (so that all individuals are allowed to interact)
could be obtained from the master equation, since in this case the
moments form a closed hierarchy. Defining the moments $M_1$ and
$M_2$ as in the Appendix, the variance $\sigma^2=M_2-M_1^2$
satisfies:
\begin{equation}
\frac{d\sigma^2}{d t}=-\sigma^2+\frac{m}{12}.
\label{eq:variance}
\end{equation}
In this limiting case (where the main feature of the model,
bounded confidence, has been lost since everybody is able to
interact with everybody), the variance reaches a steady-state in
which the width increases with noise as $\sigma\sim m^{1/2}$.

\begin{figure}
\centering
\mbox{\epsfig{file=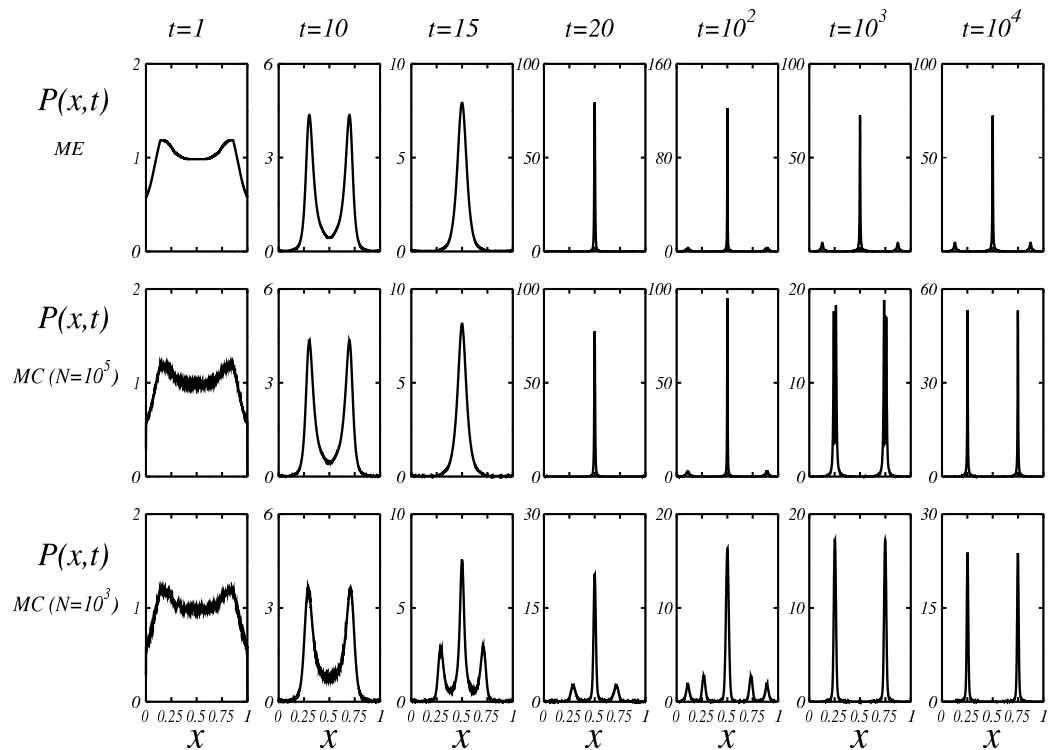,width=1.0\linewidth}}
\hfill
\caption{Probability distribution function $P(x,t)$, for
intermediate time steps, from Monte-Carlo (MC) simulations
(histograms binned with bin size $\Delta x=5\times 10^{-4}$) for two different system sizes $N=10^3$ and $N=10^4$, and the master
equation (ME) integrations of Eq.~(\ref{eq:MEN}) starting with a
flat distribution with $\epsilon=0.28$ and $m=0.01$. The distributions are an average over $10^5$ realizations for $N=10^3$ and $10^4$ realizations for $N=10^5$.
\label{fig:P_time2}}
\end{figure}

\subsection{Comparison with Monte-Carlo simulations}
\begin{figure}
\centering
\epsfig{file=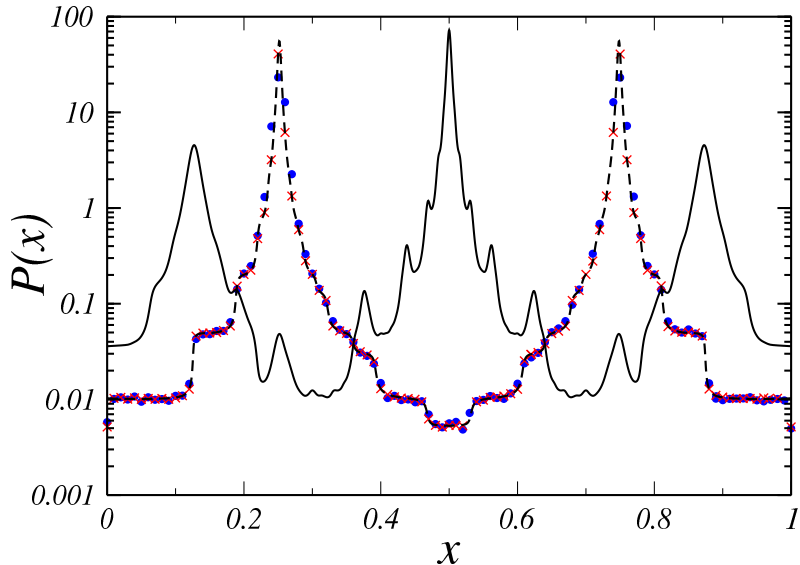,width=0.75\linewidth}
\caption{The symbols denote the steady-state probability
distribution function $ P_{st}(x)$, obtained from Monte-Carlo
simulations for system sizes $N=10^3$ ($\cdot$) and $N=10^5$
($\times$) in the case $\epsilon=0.28$, $m=0.01$. The dashed line
is obtained as an asymptotic solution of the master equation
starting with a non-uniform initial condition. These three
distributions are very similar in this logarithmic scale, although
there are differences in the height of the maxima (see Fig.
\ref{fig:P_time2}). It turns out that close to each of the two maxima, the distributions can be well fitted by a Lorentzian function\cite{ben-naim}. The solid line is the distribution $ P_{\infty}(x)$
coming from a numerical solution of the master equation
Eq.~(\ref{eq:MEN}) starting with a uniform distribution.
\label{fig:P_steady}}
\end{figure}

Once the master equation predictions have been established, and
before comparing with the results coming from the Monte-Carlo
numerical simulations using the microscopic rules, a word of
warning is, as in the noiseless case, required. In g
simulations with a finite number $N$ of individuals the dynamics of the probability distribution as well as its
asymptotic, steady-state, values might not coincide with the analysis of the master
equation. We have found this deviation to be more pronounced in
the case of being close to a bifurcation point. For example, in
Fig.~\ref{fig:P_time2} we plot the time evolution of the
probability coming from Monte-Carlo simulations of the model for
different system sizes and the results of the master equation in
the case $\epsilon=0.28$, close to a type-2 bifurcation point. It
can be seen that, although the Monte-Carlo simulation and the
master equation agree initially very well, they start to deviate
after a time that depends on the number of individuals $N$: the
larger $N$, the longer the time than the Monte-Carlo simulations
are faithfully described by the master equation. In this
particular case, $\epsilon=0.28$, it can be seen that the master
equation agrees with the Monte-Carlo simulations up to a time
$t\sim10 $ for $N=10^3$ and a time $t\sim100$ for $N=10^5$. In
view of this difference, it is surprising that the Monte
Carlo steady-state distributions show only small (although
observable) finite-size effects. Quite similar functions describe
the steady-state data for both $N=10^3$ and $N=10^5$, see
Fig.~\ref{fig:P_steady}. As can be seen in Fig.~\ref{fig:P_time2},
however, while the numerical solution of the master equation tends
to the steady-state distribution $ P_{\infty}(x)$, the Monte-Carlo
simulations tend to another distribution, $ P_{st}(x)$. These two
distributions are very different: $ P_{\infty}(x)$ has a large maximum
(large group) at $x=0.5$ and two much smaller maxima at $x\approx
0.127$ and $x\approx 0.873$, whereas $ P_{st}(x)$ has two equal maxima
at $x\approx 0.25$ and $x\approx 0.75$. Although surprising at
first, it turns out that the steady-state distribution $ P_{st}(x)$
coming from the Monte-Carlo simulations of the model is also very
close to a steady-state solution of the master equation
(\ref{eq:MEN}) having two major groups. However, this last
steady-state solution can not be obtained as an asymptotic
solution of the master equation $\lim_{t\to\infty}P(x,t)$ starting
from a uniform initial condition $P(x,t=0)=1$ for $ x\in[0,1]$. It
turns out that it is reached when starting instead from an initial
condition asymmetric with respect to the center of the interval.

Summing up, for
$\epsilon=0.28$ there are two steady-state solutions of the
master equation, $ P_{st}(x)$ and $ P_{\infty}(x)$. Starting from a uniform
initial condition, $ P_{\infty}(x)$ is the one asymptotically reached
as a solution of the master equation. However, $P_{st}(x)$ is, up to finite-size effects, the one
reached instead in the Monte-Carlo simulations. We interpret this
in terms of the relative stability of both solutions: introducing
a fluctuation $\delta P(x)$ on the solution $ P_{\infty}(x)$ it is then
possible to reach the solution $ P_{st}(x)$, but not the reverse. This
fluctuation $\delta P(x)$ needs to be asymmetric, $\delta P(x)\ne
\delta P(1-x)$ and it appears naturally in Monte-Carlo simulations
because of the finite number of individuals $N$ and it is more probable
the smaller the value of $N$. This explains that the system with
smaller $N$ deviates earlier from the solution of the master
equation. If one induces artificially\footnote{If one is not
careful enough, the perturbation might also appear as a numerical
instability of the integration method.} such a non-symmetric
perturbation in the solution $ P_{\infty}(x)$ or, alternatively, one
starts with a non-uniform, asymmetric initial condition, then the
master equation tends to $ P_{st}(x)$.

\begin{figure}
\centering
\vspace{-2.0truecm}
\epsfig{file=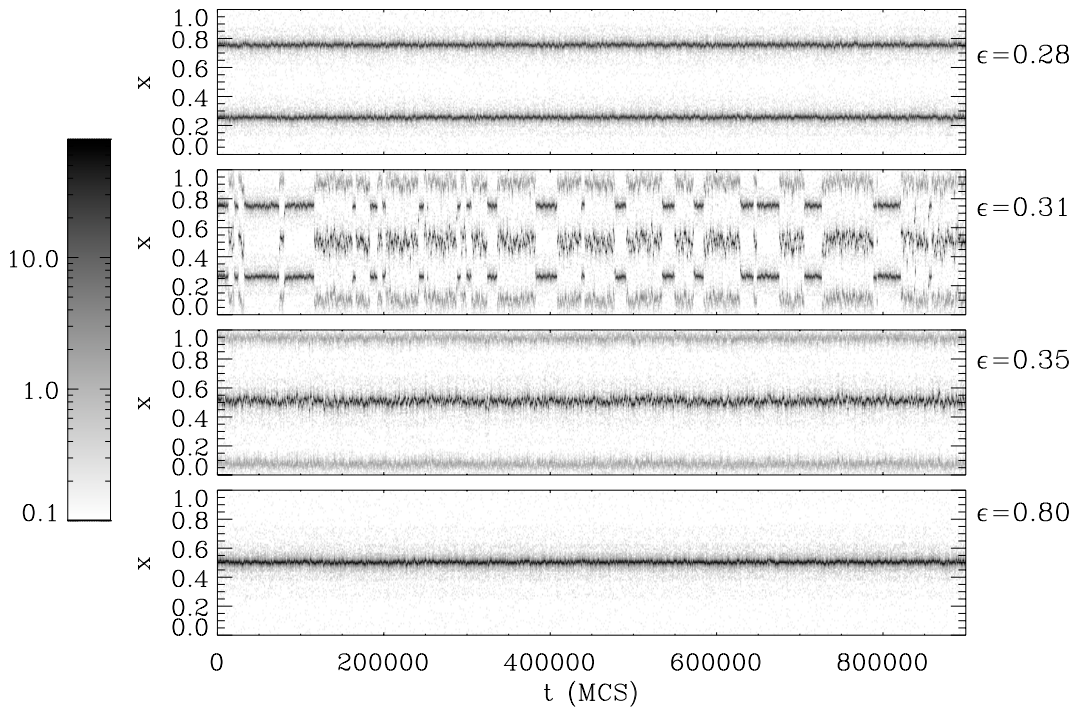,width=1.1\linewidth}
\vspace{-10.0truecm}
\caption{Time series of the opinion distributions $P(x,t)$ at four values of
$\epsilon$ and $m=0.01$ and $N=1000$. Grey scale is logarithmic,
with values smaller than 0.1 plotted in white. Simulations are
initialized with random homogeneous distributions 1000 MCS before
the first time shown in the plots. The three first panels are near
a type-2 bifurcation. At $\epsilon=0.28$ the system appears
polarized in two opinion groups. At $\epsilon=0.35$ a single major
opinion has been established, with two lateral minor groups. At
$\epsilon=0.31$ the systems fluctuates between these two states. A
single group is observed at large $\epsilon$ (bottom panel).
\label{fig:trajectories}}
\end{figure}

The existence of more that one steady solution of the master
equation seems to be a general feature. In fact, if the
distributions in Fig. \ref{fig:P_st_master} are recalculated by
slowly increasing and decreasing $\epsilon$ without resetting the
initial condition to $P(x,t=0)=1$ after each change in $\epsilon$,
we observe the hysteresis behavior typical of bistability
occurrence close to first-order transitions. Which one of the
possible steady solutions is observed in the Monte-Carlo
simulations depends on the parameter $\epsilon$. It could even
happen that the inherent fluctuations of a finite system take it
from one solution to another and back. This sort of bistability is
observed, for instance, in Monte-Carlo simulations at
$\epsilon=0.31$. As shown in Fig.~\ref{fig:trajectories}, the
evolution displays multiple jumps between two solutions: one with
two maxima of equal height and another one with a large central
maximum and two smaller maxima near the edges of the opinion
interval.

Looking at Fig.~\ref{fig:P_st_master}, one can see that for small
$\epsilon$ or, alternatively, for noise intensity $m$ larger than
a critical value $m_c$ which increases with $\epsilon$, the
bifurcations become blurred and the maxima of the distributions
are not evident, implying the inhibition of group formation. This
happens at $\epsilon\lesssim 0.1$ for $m=0.1$ and at
$\epsilon\lesssim 0.01$ for $m=0.01$. A similar effect can be
observed in the Monte-Carlo simulations and can be described in
terms of an order-disorder transition: order identified with the
state with well defined opinion groups and disorder identified
with the state without groups. To define in a
more quantitative way this transition, we have used the so-called
group coefficient $G_{M}$, which aims to characterize the
existence of groups \cite{lopez}. The definition of the group
coefficient $G_M$ starts by dividing the opinion space $[0,1]$ in
$M$ equal boxes and counting the number of individuals $l_i$
which, at time step $n$, have their opinion in the box
$[(i-1)/M,i/M]$. One next introduces an entropy-like measure
$S_M=-\sum_{i=1}^{M}\frac{l_i}{N}\ln\frac{l_i}{N}$. Note that $0
\leq S_M \leq \ln M$, and that the minimum value $S_M$ is obtained
when all the individuals are in just one box, while the maximum
value, $S_M=\ln M$, is reached when $l_{i}=N/M$, i.e. when the
opinions are uniformly distributed in the interval $[0,1]$.
Finally, the opinion group coefficient is defined as \cite{lopez}
\begin{equation}
G_M=M^{-1}\left\langle e^{\overline S_M}\right\rangle,
\end{equation}
where the over-bar denotes a temporal average in steady conditions
and $\langle \cdot\rangle$ indicates an average over different
realizations of the dynamics. Note that $1/M\le G_M\le 1$. Large
values $G_M \approx 1$ indicate that the opinions are evenly
distributed along the full opinion space (a situation identified
with disorder), while small values of $G_M$ indicate that opinions
peak around a finite set of major opinion groups (a situation
identified with order).
\begin{figure}[htp]
\centering
\mbox{\epsfig{file=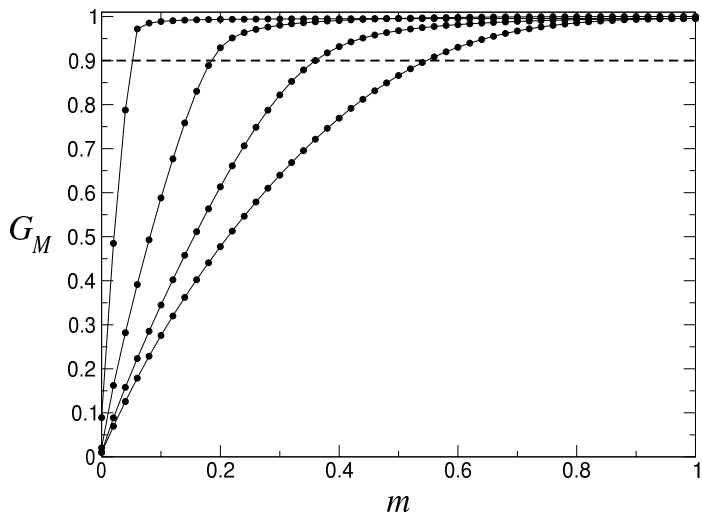,width=0.5\linewidth,clip=,angle=270}}
\hfill
\caption{Opinion group coefficient $G_{M=100}$ versus $m$ for
$\epsilon=0.05$, $0.2$, $0.4$, and $0.6$, from left to right,
respectively as obtained from Monte-Carlo simulations with
$N=10^5$ individuals (dots, the solid line is a guide to the eye).
The location of the order-disorder transition is defined as the
value $m_c$ for which $G_M=0.9$. Note that $m_c$ is an increasing
function of $\epsilon$ and that the transition is very abrupt for
small $\epsilon$. \label{groupcoefficient}}
\end{figure}

The data coming from Monte-Carlo simulations, see
Fig.~\ref{groupcoefficient}, show that $G_M$ is an increasing
function of the noise intensity $m$ and saturates to its maximum
value $G_M\approx 1$ for large enough values of $m$. The
transition from group formation to disorder will be defined,
somehow arbitrarily but precisely, as the value $m_c$ of the noise
intensity for which the group coefficient reaches the value
$G_M=0.9$. For small values of the confidence parameter $\epsilon$
the transition to the homogeneous state is abrupt and occurs for
small values of $m$. If one increases $\epsilon$, the transition
becomes less abrupt and a higher noise intensity is needed to
obtain the homogeneous, group-free, state. This last feature can
be explained by the linear stability analysis that we shall develop in
the next section.

\section{Linear stability analysis}
\label{sec:linear}
\begin{figure}
\centering
\mbox{\epsfig{file=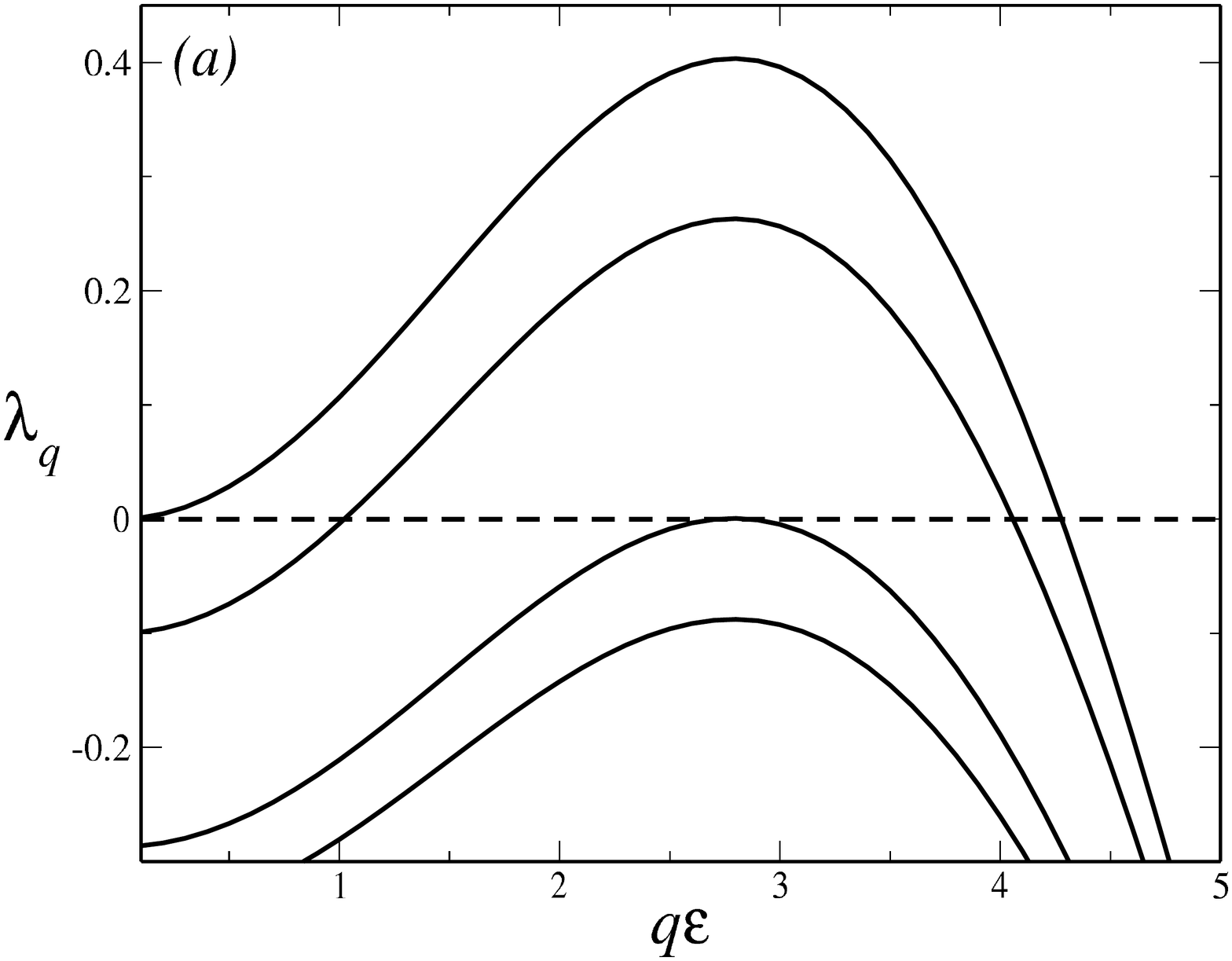 ,width=0.45\linewidth}\epsfig{file=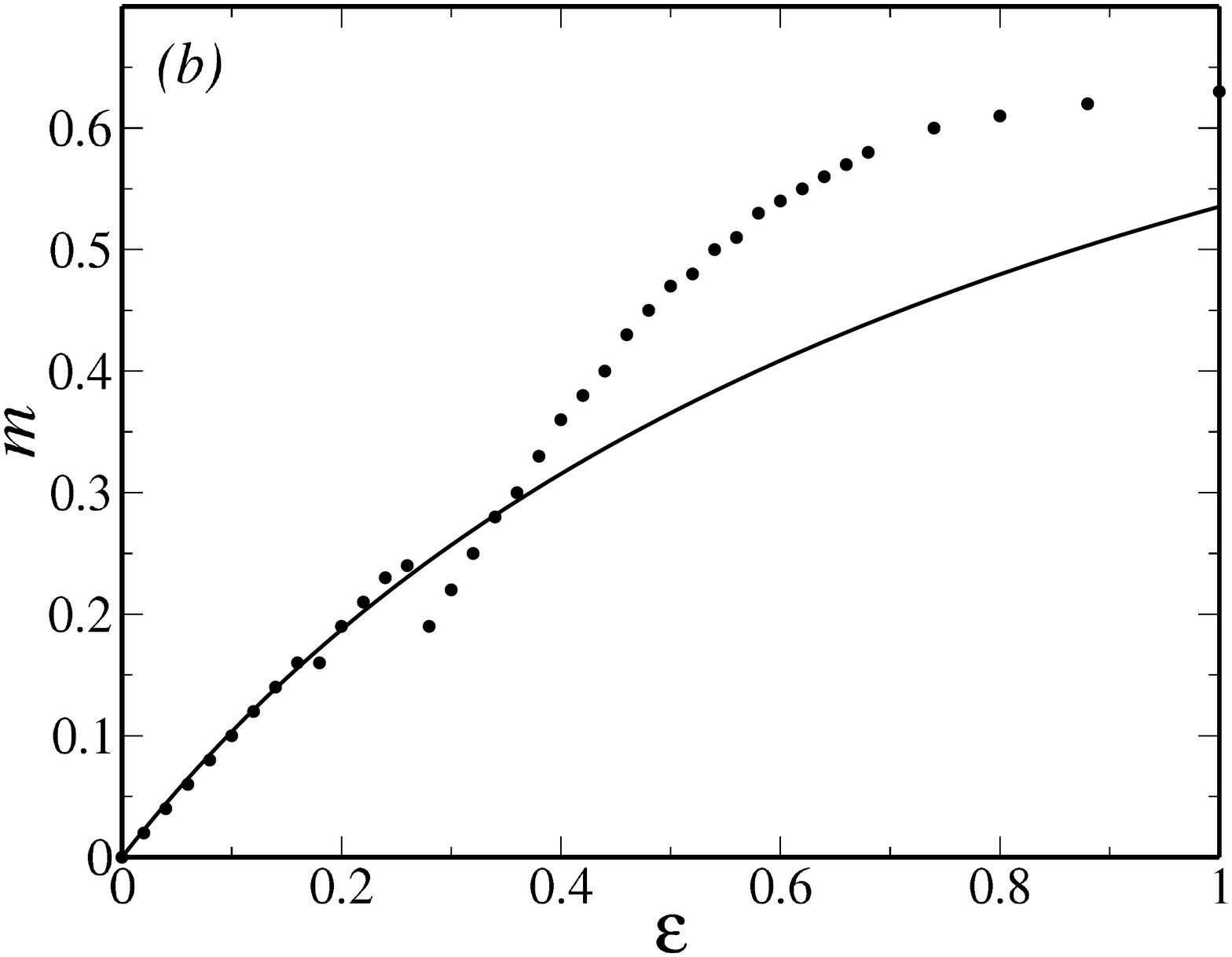 ,width=0.45\linewidth}}
\hfill
\caption{(a) Growth rate, Eq. (\ref{eq:dispersion}), of a perturbation to the
homogeneous solution as a function of $q\epsilon$ for
$\epsilon=0.35$ and $m=0.0$, $0.10$, $0.287$, and $0.35$ from top
to bottom. It shows that $\lambda_{q}$ becomes negative for high
values of $m$. (b) Phase diagram on the plane ($\epsilon,m$) as
coming from Monte-Carlo simulations with $N=10^5$ individuals
(dots), using the criterion based on the group coefficient
$G_{M=100}$ as described in the text. The solid line is the
prediction of the linear stability analysis. Groups appear below
the lines, whereas the unstructured state is stable above.
\label{fig9} }
\end{figure}

We have shown that opinion groups still form in presence of small
amounts of noise, but an unstructured state without groups
dominates the opinion space for noise larger than a critical noise
intensity $m_c$. Although the transition to group formation
is a nonlinear process, one can still derive approximate
analytical conditions for the existence of group formation in the parameter
space ($\epsilon,m$) by performing a linear stability analysis of
the unstructured solution of Eq.~(\ref{eq:ME4}). This is greatly
simplified if one neglects the influence of the boundaries and
assumes that the interval $[0,1]$ is wrapped on a circle, i.e.
there are periodic boundary conditions at the ends of the
interval. This would be a reasonable approximation to describe the
distribution far from the boundaries if $\epsilon$, which fixes
the interaction range, is sufficiently small. In this case the
homogeneous configuration $P_h(x)=1$ is an approximation to the
unstructured steady solution of the master equation. Analysis of
its stability begins by introducing
$P(x,t)=P_h(x)+A_qe^{iqx+\lambda_q t}$, where $q$ is the wave
number of the perturbation, $\lambda_q$ its growth rate and $A_q$
the amplitude. After introducing this {\sl ansatz} in
Eq.~(\ref{eq:ME4}) we find the dispersion relation giving the
growth rate of mode $q$:
\begin{equation}
\lambda_q=4\epsilon(1-m)\left[\frac{4\sin(q\epsilon/2)}{q\epsilon}-
\frac{\sin(q\epsilon)}{q\epsilon}-1\right]-m.
\label{eq:dispersion}
\end{equation}
This is plotted in Fig.~\ref{fig9}(a), for several values of $m$.
The maximum value of $\lambda_q$ occurs at
$q_{max}=2.7907/\epsilon$. It turns out that he maximum value
$\lambda_{q_{max}}$ is negative for $m>m_c$ and positive for
$\displaystyle m<m_c =\frac{\epsilon}{a+\epsilon}$ with
$a\approx0.8676$. Alternatively, for fixed $m$ the maximum growth
rate is negative for $\displaystyle\epsilon<\epsilon_c=\frac{a
m}{1-m}$, and positive for $\epsilon>\epsilon_c$. Therefore, the
homogeneous state is unstable and group formation is possible only
for $m<m_c$ or $\epsilon>\epsilon_c$. The numerical values are
$\epsilon_c=0.096$ for $m=0.1$ and $\epsilon_c=0.0088$ for
$m=0.01$, in reasonable agreement with the behavior observed for
the master equation dynamics in Fig. \ref{fig:P_st_master}.
Comparison with Monte-Carlo simulations is performed in
Fig.~\ref{fig9}(b) where we plot the critical value $m_c$ obtained from
the group coefficient $G_M$ as described earlier. We see in the
figure that the agreement is very good for small $\epsilon$ but
deviates for larger values. This is consistent with the fact that
neglecting boundary effects is expected to be appropriate only for
small $\epsilon$. Finally, it is possible to estimate roughly the
number of groups $n$ by a simple argument: $n$ is related to the
wavelength of the maximum growth as $n=q_{max}/2\pi$ or
$n=0.444/\epsilon$ for $\epsilon>\epsilon_c$. This result is in qualitative agreement with
the  $\frac{1}{2\epsilon}$-rule, which says that the number
of major groups after group formation is roughly determined as the
integer part of $\frac{1}{2\epsilon}$ (see \cite{lorenz1,lorenz2} for
details).

\section{Summary and conclusions}
\label{sec:conclusions}

In this paper we have studied the Deffuant {\sl et al.} model for
continuous--opinion dynamics under the presence of noise. Besides
the usual rules of the model, we give each individual the
opportunity to change, with a certain probability $m$, his opinion to a
randomly selected opinion inside the whole opinion space. The final
behavior depends of the confidence or interaction parameter
$\epsilon$ and the noise intensity $m$.

We have first reviewed the original noiseless version of the model.
We have shown that for small number of individuals $N$ and
depending of the particular realizations, the exact location of
the opinion groups might vary with respect to the predictions of
the master equation. In particular, minor groups might not appear
and there could be realizations in which even the number of
observed major groups differs from the prediction of the master
equation.

We have derived (Appendix) a master equation for the probability
density function $P(x,t)$ which determines the individuals density
or distribution in the opinion space. Numerical integration of
this equation from uniform initial conditions reveals that for
$m>0$ the steady distributions are no longer delta-functions as
for $m=0$, but are still peaked around some well defined maximum
value, with some non-vanishing width. By looking at those maxima
we are able to identify the same type of bifurcations than in the
noiseless case \cite{redner,lorenz1}. At variance with the
noiseless case, the location of the maxima (the central opinion of
the groups) does not depend on $\epsilon$ until a new bifurcation
point is reached.

We have also found that in the noisy case the asymptotic
steady-state probability distributions reached by Monte-Carlo
simulations might not coincide with the ones obtained from the
master equation dynamics starting from the same symmetric initial
condition. This deviation is more pronounced in the case of being
close to a bifurcation point. In particular, we have presented a
situation where, starting from a uniform initial condition, a
particular stationary distribution, $ P_{\infty}(x)$, is actually
reached by the master equation but another distribution , $ P_{st}(x)$, is the one
reached instead in Monte-Carlo simulations. The time in which the
Monte-Carlo simulations begin to deviate from the master equation
depends on system size: the smaller the size, the earlier the
deviation occurs although the final Monte-Carlo distribution
$ P_{st}(x)$ shows only small size effects. Remarkably, $ P_{st}(x)$ turns
out to be close to another steady solution of the master equation.
Thus, the discrepancy observed during the dynamics does not seem
to be simply a trivial finite-$N$ effect. We interpret it in terms
of the relative stability of both solutions by adding an
asymmetric perturbation to $ P_{\infty}(x)$. It is then possible to
reach the solution $P_{st}(x)$ obtained from Monte-Carlo simulations, but not the other way around . Asymmetric
fluctuations appear naturally in Monte-Carlo dynamics because of
the finite number of individuals, and are larger for smaller $N$. We
have also shown that the fluctuations present in Monte-Carlo
simulations are even able to induce jumps from one solution to
another and back.

An order-disorder transition to group formation induced by noise
has been characterized using the so-called group coefficient
$G_{M}$ for the simulations performed with Monte-Carlo dynamics.
We have found that $G_M$ is an increasing function of the noise
intensity $m$ and saturates to its maximum value $G_M\approx 1$
for large enough values of $m$. For small values of the confidence
parameter $\epsilon$ the transition to the disordered state is
abrupt and occurs for small values of $m$. If one increases
$\epsilon$, the transition becomes less abrupt and a higher noise
intensity is needed to obtain this state. Using a linear stability
analysis of the unstructured (no groups) solution of the master
equation we have derived approximate conditions for opinion group
formation as a function of the relevant parameters of the system.
We have found qualitative agreement between the linear stability
analysis and numerical simulations. The agreement is better for
small values of $\epsilon$ where boundary effects, neglected to
make feasible the linear analysis, are less important. However, we
should emphasize that the pattern selection of this model is, with
noise and without it, intrinsically a nonlinear phenomenon and
obtaining the exact critical conditions for opinion group
formation remains a challenge.

Our work stresses the importance that fluctuations and finite-size
effects have in the dynamics of social systems for which the
thermodynamic limit is not justified \cite{tt2007}. Further work
will address the effect that these ingredients have in the
dynamics of continuous--opinion models in the presence of an
external influence, or forcing, representing the role of
advertising.

\section*{Acknowledgments}
We are grateful to V. M. Egu\'{\i}luz and Pere Colet for interesting discussions.
We acknowledge the financial support of project FIS2007-60327 from
MICINN (Spain) and FEDER (EU) and project FP6-2005-NEST-Path-043268 (EU).

\appendix
\section*{Appendix}
\setcounter{section}{1}

Here we derive the master equation, i.e. the evolution equation
for $P_{n}(x)$, the probability density function (pdf) of the opinions at step $n$ for the model introduced in this paper. Note that $P_n(x)$ is constructed from the
histogram of all individual opinions $x^i_n$ at step $n$. Let us
first find the evolution of the pdf for those two particular
individuals $i,j$ that have been selected for updating at step $n$
according to the basic rule Eq.~(\ref{eq:rule}). We will denote by
$P^{i}_{n+1}(x)$ the pdf of the opinion of individual $i$ at the
step $n+1$, i.e the probability that $x^i_{n+1}$ adopts the value
$x$. According to that rule, it is straightforward to derive the
evolution equation
\begin{eqnarray}
\label{eq:ME1}
P^{i}_{n+1}(x)&=
\int_{|x^{i}_n-x^{j}_n|<\epsilon}dx^{i}_ndx^{j}_nP_{n}(x^i_n)P_{n}(x^j_n)\delta
\left( x-\frac{x^i_n+x^j_n}{2} \right)
\nonumber\\
&
+\int_{|x^{i}_n-x^{j}_n|>\epsilon}dx^{i}_ndx^{j}_nP_{n}(x^i_n)P_{n}(x^j_n)\delta(x-x^{i}_n),
\end{eqnarray}
and a similar expression for $P^{j}_{n+1}(x)$. The integrals over
$x{^i}_n$ and $x{^j}_n$ run both over the interval $[0,1]$. In
this equation an independence approximation for the variables
$x^{i}_n,x^{j}_n$ has been implicitly assumed, i.e. their joint
pdf $P_{n}(x^{i}_n,x^{j}_n)$ is supposed to factorize as
$P_{n}(x^{i}_n,x^{j}_n)=P_{n}(x^{i}_n)P_{n}(x^{j}_n)$. 
This is an uncontrolled approximation whose validity can only be established by an ulterior comparison with the Monte-Carlo simulation of the microscopic rules.
For the individuals $k\ne i,j$ whose opinion does
not change at time step $n+1$ we have simply
$P^k_{n+1}(x)=P^k_{n}(x)$.

The pdf $P_{n+1}(x)$ has several contributions: (i) With
probability $1-m$ two individuals, say $i,j$, are chosen for
updating according to the basic evolution rule Eq.~(\ref{eq:rule})
and $N-2$ variables remain unchanged. (ii) With probability $m$
one individual is chosen for updating according to the noise rule
and $N-1$ variables remain unchanged; the new opinion of the selected individual is sampled from an, in principle, arbitrary distribution
$P_a(x)$, although in this paper we have taken throughout that $P_a(x)$ is the uniform distribution $P_h(x)=1$ in the interval $[0,1]$. After
consideration of these contributions we are led to the
evolution equation
\begin{eqnarray}
\label{eq:ME2}
P_{n+1}(x)&=(1-m)\left[\frac{N-2}{N}P_{n}(x)+\frac{1}{N}P^{i}_{n+1}(x)+\frac{1}{N}P^{j}_{n+1}(x)\right]\nonumber\\
& +m\left[\frac{N-1}{N}P_{n}(x)+\frac{1}{N}P_a(x)\right ].
\end{eqnarray}
Replacing $P^{i}_{n+1}(x)$ and $P^{j}_{n+1}(x)$ from
Eq.~(\ref{eq:ME1}) one obtains after some algebra
\begin{eqnarray}
\label{eq:ME3}
P_{n+1}(x)&=P_{n}(x)+\frac{(1-m)}{N}\left[4\int_{|x-x'|<\epsilon/2}dx'P_n(2x-x')P_n(x')\right.\nonumber\\
&
\left.-2P_n(x)\int_{|x-x'|<\epsilon}dx'P_n(x')\right]+\frac{m}{N}\left[P_a(x)-P_n(x)\right].
\end{eqnarray}
The integrals over $x'$ run over the interval $[0,1]$, and it has
to be imposed that $P_n(x)=0$ if $x \notin [0,1]$. We now take the
continuum limit $P_{n}(x)\rightarrow P(x,t)$ with a time
$t=n\Delta t$ and taking the limit $\Delta t=1/N\rightarrow 0$ as
$N\rightarrow \infty$, to obtain:
\begin{eqnarray}
\label{eq:ME4}
\frac{\partial P(x,t)}{\partial t}&=(1-m)\left[4\int_{|x-x'|<\epsilon/2}dx'P(2x-x',t)P(x',t)\right. \nonumber\\
&
\left.-2P(x,t)\int_{|x-x'|<\epsilon}dx'P(x',t)\right]+m\left[P_a(x)-P(x,t)\right],
\label{eq:MEN}
\end{eqnarray}
which is the master equation of the Deffuant {\sl et al.} model in
the presence of noise and the basis of our analysis.
The noiseless case, $m=0$, was first obtained in reference
\cite{redner}. We note here the symmetry property of the master
equation: if the initial condition is symmetric around the central
point $x=1/2$, namely that $P(x,t=0)=P(1-x,t=0)$, then this
property holds for any later time, $P(x,t)=P(1-x,t),\forall t>0$.

The time evolution of the first moments of $P(x,t)$ can be
computed from the master equation. Defining the moments as
$M_k(t)=\int dx x^kP(x,t)$ one finds easily that $\displaystyle
\frac{dM_0}{dt}=0$ (normalization condition) and that the first
moment evolves as $\displaystyle \frac{dM_1}{d
t}=m\left(M_1^a-M_1\right)$, being $M_1^a$ the first moment of the distribution $P_a(x)$. Therefore, if $m>0$, the
average opinion tends to $M_1=M_1^a$ independently of the initial
condition, and it is always conserved in the noiseless case $m=0$.
Expressions for higher-order moments can only be obtained in the special
case $\epsilon \ge 1$, as discussed in the main text.

\end{document}